# Fabrication of diamond diffraction gratings for experiments with intense hard x-rays

M. Makita [a], P. Karvinen [a], V. A. Guzenko [a], N. Kujala [b], P. Vagovic [c], and C. David [a]

*(a) Paul Scherrer Institut, Villigen PSI, Switzerland*
*(b) European XFEL GmbH, Holzkoppel 4, 22869 Schenefeld, Germany*
*(c) CFEL, Deutsches Elektronen-Synchrotron, Notkestrasse 85, Hamburg, Germany*

**Abstract**

The demands on optical components to tolerate high radiation dose and manipulate hard x-ray beams that can fit the experiment requirements, are constantly increasing due to the advancements in the available x-ray sources. Here we have successfully fabricated the transmission type gratings using diamond, with structure sizes ranging from few tens of nanometres up to micrometres, and aspect ratio of up to 20. The efficiencies of the gratings were measured over a wide range of photon energies and their radiation tolerance was confirmed using the most intense x-ray source in the world. The fidelity of these grating structures was confirmed by the quality of the measured experimental results.

1. Introduction

The modern bright x-ray sources offer significant contributions to the research interest across the scientific fields, such as material physics, biomedical science and high energy density research [1 – 3]. More and more facilities have gone through significant progresses in order to provide brighter and more coherent x-rays, one of such advancement is called X-ray Free Electron Lasers (XFEL) which are either already in operation or about to [4 – 6]. XFELs offer intense, coherent femto-second pulses, resulting in characteristic peak brilliance values a billion times higher than that from conventional synchrotron facilities [4]. Such pulses result in extreme peak radiation levels on the order of several Terawatts per square centimetre for any optical component in the beam, and can reach ablation threshold of many materials. It is, therefore, of key importance to fabricate optics that are robust enough to withstand such radiation levels, and yet are capable of efficiently manipulating these x-ray beams for experimental purposes. These problems could be circumvented either by using novel materials or introducing unusual designs. Indeed several optical elements were fabricated having such concepts in mind, for example from Silicon [7], or from Tungsten [8]. However, the intrinsic properties of these materials, such as the radiation tolerance and x-ray transmission capability, limit their applicability. Among various materials available the currently ideal material for hard x-ray application is diamond, which is known for its high thermal conductivity, thermal stability and low x-ray absorption coefficient. These properties allow for robust optics to be developed using diamond structures that could be obtained combining electron-beam lithography and inductively coupled plasma (ICP) assisted reactive ion etching (RIE). We present here the development of diamond x-ray optical elements in the form of transmission-type diffraction gratings with structure widths ranging from nanometres to micrometres, and etched depths of up to 9 μm and sidewall slopes of < 5 degree.

## 2. Material and Methods

Polycrystalline diamond membranes (Diamond Materials, Freiburg), made by chemical vapour deposition (CVD), were used as a material for the optics fabrication. The dimensions of the membranes vary between 1 mm and 5 mm diameter, all of them are 10 μm thick and supported by a 500 μm thick Si frame.

The fabrication process started with the conditioning of the diamond membranes substrates by a cleaning procedure including a Piranha bath and rinsing with 10 % Hydrofluoric acid, followed by oxygen plasma treatments. These processes steps are crucial for removing particles contaminating the surface and to promote adhesion of the mask layers in the following steps. The diamond substrates were then coated with 15 nm Cr, followed by another oxygen plasma treatment. This procedure has two purposes: (i) Oxygen plasma treated Cr serves as an adhesive layer for the used Hydrogen silsesquioxane (HSQ) electron-beam resist layer, and (ii) as a conductive layer to avoid excessive charging of the electrically insulating diamond surface during electron-beam exposure. A negative-tone resist HSQ (FOx-16™, Dow Corning) was spin-coated at 6000 rpm for 60 s on the Cr-coated membranes, yielding ~ 450 – 500 nm thick resist layer. In case of structures smaller than ~40 nm, a Methyl Isobutyl Ketone (MIBK) diluted HSQ (1:1 by volume) was used, and resulted in a thickness of at the same coating parameters. The grating patterns were exposed onto the resist by electron-beam lithography at 100 keV electron energy using a Vistec EBPG 5000+ES writer, either at 2 – 10 nA using a 200 um final aperture for grating structures below 60 nm, or at 30 – 150 nA using 400 μm aperture for larger structures. The optimal exposure parameters were found by test exposure series. We found optimal exposure doses of between 5000 μC/cm$^2$ and 13000 μC/cm$^2$, depending on the grating line widths and the HSQ resist shelf time, which seem to result in significantly varying sensitivity. Moreover, its adhesion generally became weaker for smaller patterns, underexposed patterns typically result in structures that are not fully developed and are prone to detach, and hence dose varying for each pattern was necessary for optimisation. For diamond membranes, which typically exhibit a curved surface, the surface height measurements for e-beam writing fields were fitted with polynomial function to compensate for its profile.

The development was performed by immersion of the exposed chips in a mixture of Sodium hydroxide (NaOH) and water (1:3 by volume), for 5-6 minutes, followed by rinsing in de-ionized water, and in isopropanol. For structure widths above 50 nm, the chips can be dried in a N$_2$ gas jet. For structures smaller than 50 nm, the chip was dried using critical point drying to avoid structure collapse due to the capillary force.

After the development, the samples were etched in a Cl$_2$/O$_2$ plasma to transfer the HSQ pattern into a Cr layer, revealing the underlying diamond. The selectivity of this etch process is relatively high, however too long treatment results in a slightly reduced HSQ mask layer. The samples are then baked at 300 °C for 40 minutes. This process significantly hardens the HSQ mask, therefore enhancing the etching selectivity in the subsequent pattern transfer into the diamond. The diamond is etched by oxygen ICP-RIE (OXFORD PlasmaPro100). The used recipe was optimized for anisotropy and selectivity, by tuning the ion energy and ion density in the ICP etcher. The process typically requires readjustments of these parameters after ~1-2 μm depth of etching, in order to minimise the sidewall slope.

## 3. Results & Discussion

Fig. 1 (a), (c) shows a side-view of a linear transmission grating, with a size of 2 mm by 2 mm, a pitch of 200 nm, and an etched depth of ~1.4 μm. The typical aspect ratios of the lines of these gratings are 10 – 20, depending on the duty cycle. Since the exposures were carried out on thin membranes consist of Carbon only, the proximity effect by back-scattered electrons was negligible.

Initial trials showed that narrow (~ 100 nm or below) and long (~ 1 – 2 mm) HSQ structures collapsed during the diamond etching step, presumably due to the severe heating and sputtering when going through the ICP process. To improve the structure stability during the ICP etching the grating lines were connected by support structures, where the developed HSQ pattern resembled an overlaid mesh of narrow line and wide pitch. The x-ray optical effect of these support structures turned out to be negligible, as the duty cycle of the support structures was only of the order of 0.05.

These nanoscale pitch gratings were used as a hard x-ray beam splitter to divert a small portion of the beam for quality analysis or sample probing purposes. Hence the diffraction efficiencies and the structure homogeneities of the gratings are of crucial parameters to better understand and optimise the fabrication processes. We have tested several gratings in the hard x-ray range (6 – 18 keV photon energy) at synchrotron facility PETRA III, Deutsches Elektronen-Synchrotron (DESY).  Fig. 2a shows the efficiency map of a typical grating over its entire surface of 2 mm by 2 mm indicating the homogeneity of the grating structures. The variation of the efficiency across the grating lies within 30%. A faint grids shaped inhomogeneity can be observed in the image. The pitch of this grid corresponds to the stitching field size of the electron-beam exposure. This is mainly due to the curvature of the diamond membranes, resulting in small errors of the focus settings and e-beam height-mapping mismatch. This has an effect of the electron-beam size and writing fields stitching accuracy, and causes local variations of the grating line width and depth.

The diffraction efficiency dependency on the x-ray energy for two of the gratings is shown in fig. 2b. The measured efficiency reaches approximately 55 % (sample 1) and 35 % (sample 2) of the theoretically calculated values [9]. This discrepancy in efficiency values between the theory and the experimental value mainly arises from the fact that, in theoretical calculations the structures consist of perfect binary square structures and the imperfections arising from the fabrication steps are not taken into consideration. The fluctuations in efficiency evaluation are caused by fluctuations of the experimental conditions, such as the beam parameters, the detector resolution and sensitivities.

We have also tested gratings of various pitches ranging between 40 nm and 200 nm at an XFEL facility, Linac Coherent Light Source (LCLS) at the Stanford National Accelerator Laboratory, for x-ray energies between 4 keV and 8 keV. Optical examinations of these gratings after the experiment confirmed that they have proven to tolerate the extreme radiation doses of one of the most intense x-ray laser source in the world. More details about the performance of the gratings and experiment details are explained elsewhere [10 – 12].

In addition to nanometre pitched beam-splitter gratings, the fabrication method can be applied to produce a variety of larger and deeper etched structures. Fig. 3 shows micrometre-sized checker board (fig. 3a), pillar (fig. 3b) and line (fig. 3c) structures etched up to ~ 7 μm depth. Using an HSQ layer of 600 – 700 nm, and hard-baking for over 60 minutes, allows for a longer lasting HSQ etch mask, and hence deeper etched structure. As such, the limit of mask layer selectivity we have achieved so far was up to ~13, resulting in etched depth of 8 – 9 μm. The slope angles of the sidewalls were measured 5 degrees or below, using SEM with the samples tilted to 85 degree. The maximum aspect ratio of micron-sized structures was on the order of 10. As these relatively large structures are rigid enough, support structures were not necessary.

Although structures with line width < 200 nm can be fabricated with relatively straight walls up to an aspect ratios of ~ 10, the fabrication of sub-micron structures with straight walls with aspect ratio > 2 is a challenge due to the process of the etching. In fig. 1 (a) indeed the bottom of the gratings shows slopes and tapered structures. This effect could be minimised by adjusting the applied power during the etching process, e.g. by increasing the applied power for ion energy (RF power) and slightly reduced power (ICP power) for density control. This tuning seems to result in 'opening up' the sloped wall laterally by physical sputtering but without affecting much the etched depth. The result of this etching parameter tuning is shown in Fig. 3. Typically the structures were etched more than 5 μm in depth. For every 1.5 μm vertical etching the applied powers were adjusted to straighten the wall before continuing to etch further in vertical direction. Some change in sidewall slope can be seen in the pillar structures, which is presumably due to erosion of the resist layer. Similar phenomena was also seen using Al etch masks [13].

These devices consisting of well-defined steep walled structures exhibit excellent phase shifting capabilities for hard x-rays without introducing significant phase errors. The gratings have been tested at a synchrotron (PETRA III P10 beamline at the GINNIX setup of Göttingen University) using 8 keV photon energies for the wave-front metrology of KB and waveguide beams. The aim was to develop single-shot capable wavefront sensing using single 2D phase grating applicable for pulsed X-ray sources. Fig. 4 shows the interferogram (self-image) recorded at the detector plane generated by single checkerboard type diamond grating. The amplitude and the phase have been successfully reconstructed from only single interferogram (fig. 4 (b, c)). The detailed description of the experimental method and the results are to be submitted for publication [14].

## 4. Conclusion

We have optimised exposure and etching parameters of nano- and micro- structure into diamond membranes using HSQ resist. Using 100 keV electron-beam lithography and ICP-RIE etching we fabricated high quality linear gratings with pitches varying from ~40 nm to 4 μm, and aspect ratios up to 20. Side wall sloping for micron sized structures was minimised by tuning the ICP applied power during the etching process, resulting in a side wall slope of less than < 5 degree over 9 μm height structures. The gratings were tested under extreme x-ray intensity and have proven their efficiency as well as tolerance to the radiation dose. These characteristics make them suitable for applications in hard x-ray beam characterisation, bio-medical imaging and x-ray sample probing at intense synchrotron and XFEL facilities.

**Acknowledgments**

The x-ray experiments shown here were performed at the P10 beamline, PETRA III at DESY, Hamburg, and at the XPP and CXI beamlines at the LCLS, SLAC Stanford National Accelerator Laboratory, California. The authors thank the beamline staff of these facilities for their support.

The research leading to these results has received funding from the European Community's Seventh Framework Programme (FP7/2007-2013) under grant agreement no. 290605 (PSI-FELLOW/COFUND), and from the EU-H2020 Research and Innovation programme under grant agreement No 654360 NFFA-Europe.

**References**

[1] S. K. Sundaram et al., "Inducing and probing non-thermal transitions in semiconductors using femtosecond laser pulses", Nat. Mater. 1, 217 (2002)

[2] K. Nagaya et al., "Ultrafast Dynamics of a Nucleobase Analogue Illuminated by a Short Intense X-ray Free Electron Laser Pulse", Phys. Rev. X. 6, DOI: 10.1103/PhysRevX.6.021035, (2016)

[3] L. B. Fletcher et al., "Ultrabright X-ray laser scattering for dynamic warm dense matter physics", Nat. Phot. 9, 274-279 (2015)

[4] P. Emma et al., "First lasing and operation of an angstrom-wavelength free-electron laser", Nat. Phot. 4, 641-647 (2010)

[5] N. E. Ipe, et al., "The Linac Coherent Light Source", Proceedings of SATIF5 (Specialists Meeting on Shielding Aspects of Accelerators, Targets and Irradiation Facilities), Paris, July 2000.

[6] M. Yabashi et al., "Overview of the SACLA facility", J. Sync. Rad. 22, 477-484, (2015)

[7] S. Rutishauser et al., "Exploring the wavefront of hard x-ray free-electron laser radiation", Nat. Comms. 3, 947 (2012)

[8] D. Nilsson et al., "Thermal stability of tungsten zone plate for focusing hard x-ray free-electron laser radiation", New. J. Phys. 14, 043010 (2012)

[9] X-ray Interactions with Matter, c1995-2010: The Centre for x-ray optics, http://henke.lbl.gov/optical_constants/tgrat2.html

[10] C. David et al., "Nanofocusing of hard X-ray free electron laser pulses using diamond based Fresnel zone plates", Sci. Rep. 1:57, DOI: 10.1038/srep07644, (2011)

[11] P. Karvinen et al., "Single-shot analysis of hard x-ray laser radiation using a noninvasive grating spectrometer", Opt. Lett. 37, 5073-5075 (2012)

[12] M. Makita et al., "High-resolution single-shot spectral monitoring of hard x-ray free-electron laser radiation", Optica 2, 10, 912-916 (2015)

[13] B. Nöhammer et al., "Diamond planar refractive lenses for third- and fourth-generation X-ray sources", J. Sync. Rad. 10, 168-171 (2003)

[14] P. Vagovic et al., "Characterisation of the X-ray wavefront using single two-dimensional phase grating interferometry" (in preparation)

[15] M. Takeda et al., "Fourier-transform method of fringe-pattern analysis for computer-based topography and interferometry", J. Opt. Soc. Am., OSA 72, 156-160 (1982)

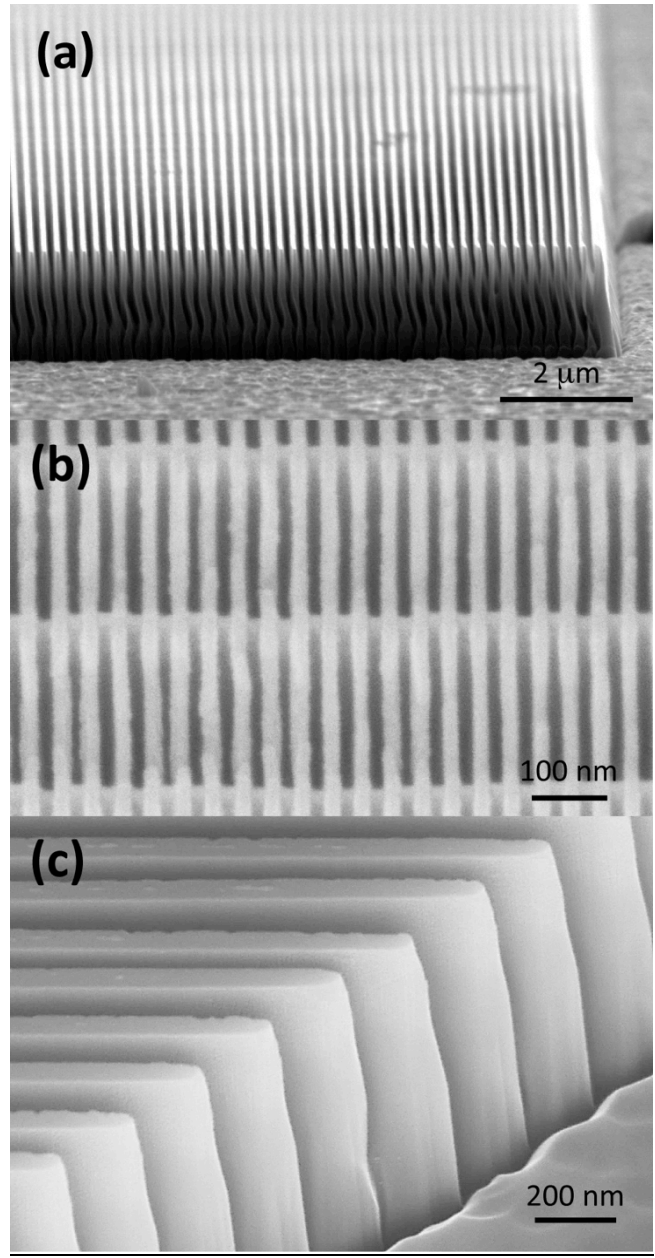

Figure 1: (a) Side view of a linear grating with a pitch of 200 nm, and a height 1.4 µm. (b) Top view of a linear grating with a pitch of 43 nm with support structures to stabilize the grating. (c) Side-wall view of grating lines with 200 nm pitch, aspect ratio 20, after tuning the ICP etching parameters.

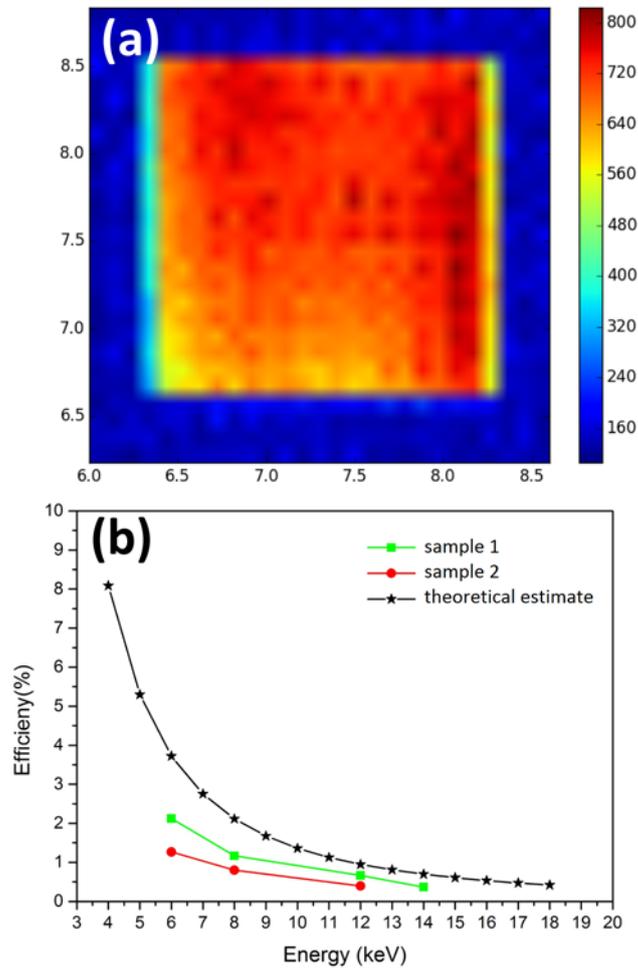

Figure 2: (a) X-ray diffraction efficiency map of a 200 nm pitch grating measured with 12 keV x-rays. Values are arbitrary units. The faint grid-like patterns seen across the grating are writing field borders. (b) Grating diffraction efficiency dependence on x-ray energy. Two sample gratings (red and green curves) of 200 nm pitch were experimentally measured and compared with theoretical limit (black).

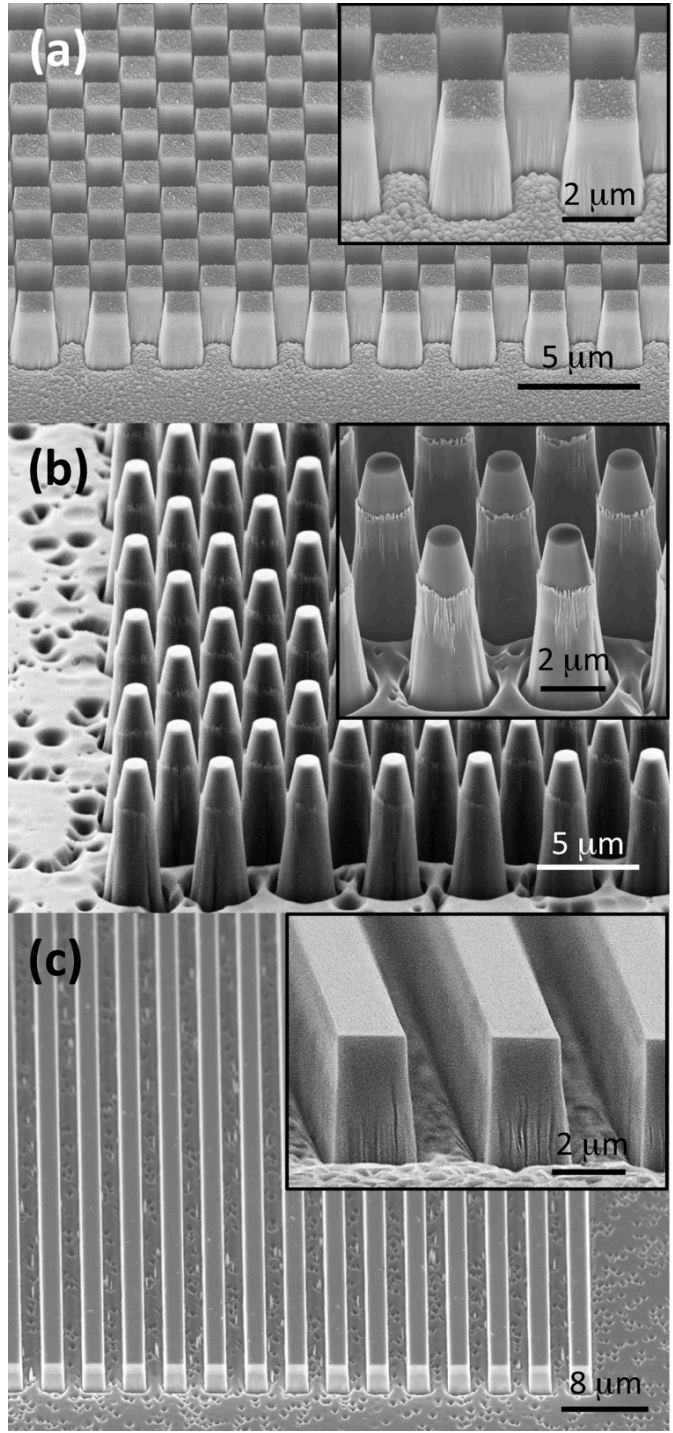

Figure 3: Micron-sized x-ray phase shifter gratings. (a) checker-board, (b) circular pillars, and (c) line patterns. Samples were tilted either 45 degrees (fig. (a) and (b)) or 85 degrees (zoomed image in fig. (c)). The ICP etching parameters were tuned to minimise the sidewall slopes.

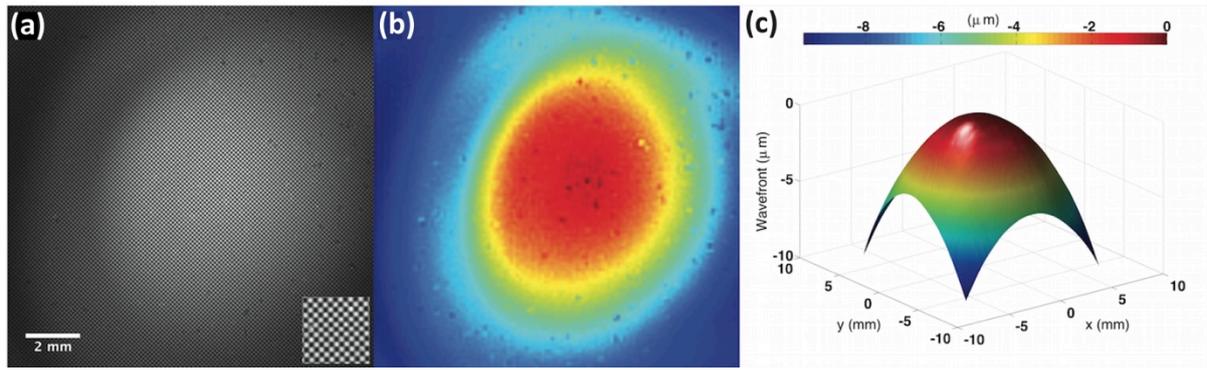

Figure 4: (a) Measurement of the geometrically magnified self-image recorded at the detector plane at the distance 5 m from the X-ray wave-guide. The inset in lower right corner is the zoomed central area of the image. The checkerboard grating with pitch of 4 µm was placed ~126 mm downstream the waveguide resulting in 40-fold magnification. Figure (b) and (c) are the demodulated amplitude and the phase of the x-ray wavefront, respectively, at the detector plane using Fourier method [15].